\documentclass[preprint,nofootinbib,showpacs,titlepage,tightenlines]{revtex4}

\usepackage{graphicx}

\newcommand{\lsim}{\raisebox{0.3mm}{\em $\, <$} 
\hspace{-3.3mm} \raisebox{-1.8mm}{\em $\sim \,$}}

\def\mbb{\langle m \rangle_{\beta\beta}}
\def\ma{{\mbox{\scriptsize{max}}}}
\def\mi{{\mbox{\scriptsize{min}}}}
\def\atm{{\mbox{\scriptsize{atm}}}}
\def\CH{{\mbox{\scriptsize{CH}}}}
\def\eV{\mbox{eV}}

\begin{document}

\preprint{hep-ph/0212240}

\title{
Double Beta Decay Constraints on Neutrino Masses and Mixing; 
Reanalysis with KamLAND Data
}

\author{Hisakazu Minakata}
\email{E-mail: minakata@phys.metro-u.ac.jp}
\author{Hiroaki Sugiyama}
\email{E-mail: hiroaki@phys.metro-u.ac.jp}
\affiliation{Department of Physics, Tokyo Metropolitan University,
Hachioji, Tokyo 192-0397, Japan}

\date{\today}

\begin{abstract}
In the light of KamLAND data just released, we reanalyze and 
update the constraints on neutrino masses and mixing parameters, 
the most general ones that can be derived in three-flavor mixing 
scheme of neutrinos with use of the bounds imposed by neutrinoless 
double beta decay search and reactor experiments.
We point out that with KamLAND data and assuming Majorana neutrinos 
one can derive, for the first time, 
an upper bound on neutrino contribution to the cosmological $\Omega$ 
parameter, $\Omega_{\nu} \leq 0.070 h^{-2}$, 
by using the current upper bound on mass parameters obtained by 
Heidelberg-Moscow group, 
$\langle m \rangle_{\beta \beta} < 0.35$ eV (90 \% CL).
If the bound is tighten to 
$\langle m \rangle_{\beta \beta} < 0.1$ eV 
by future experiments, it would lead to a far severer bound 
$\Omega_{\nu} \lsim 0.01 h^{-2}$. 
\end{abstract}


\maketitle

\section{Introduction}

The solar neutrino problem is finally solved by the KamLAND 
experiment \cite {KamLAND}, which observed a clear deficit 
of about 40 \% in $\bar{\nu}_e$ flux from reactors located 
in all over Japan. 
It pin-points the Large-Mixing-Angle (LMA) region as the unique 
solution of the solar neutrino problem based on the MSW matter 
enhanced neutrino flavor transformation \cite{MSW}, excluding 
the remaining LOW and the vacuum (VAC) solutions. 
While the LMA solution was preferred over others on statistical 
ground in combined analyses of solar neutrino data \cite{solar} 
it is by no means true that it was selected out as the unique 
solution. It is because the values of $\chi^2$ of the LOW and 
the VAC solutions, $\sim$ unity per degrees
of freedom\cite{solaranalysis}, were acceptable
and hence it was not possible with only solar neutrino data 
to uniquely idetify the LMA solution. 

It implies, first of all, that the three-flavor mixing scheme 
of neutrinos had further support by the experiments, and 
the structure of lepton flavor mixing is established 
together with the existing evidences which come 
from the atmospheric \cite{SKatm} 
and the long-baseline accelerator experiments \cite{K2K}. 
Given our above understanding of the KamLAND result it also 
means that the experiment gave the first direct proof of 
(almost) pure vacuum neutrino oscillation, the 
fascinating phenomenon first predicted by 
Maki, Nakagawa, and Sakata \cite{MNS}, 
and first applied to the solar neutrino problem by Pontecorvo 
\cite{pontecorvo}.

The resolution of the solar neutrino problem implies that 
we now know within certain uncertainties the values of 
$\Delta m^2_{12}$ and $\sin^2{2\theta_{12}}$. 
It is an urgent task to uncover the all possible implications 
of the far more constrained values of these mixing parameters. 
We address in this paper the impact of the KamLAND result 
on masses and possible mass patterns of neutrinos.

We re-analyze the constraints on neutrino masses and mixing 
parameters imposed by neutrinoless double beta decay 
experiments in the light of the first results of KamLAND 
experiment under the assumption of Majorana neutrinos.  
See ref.~\cite{vogel} for the present status of the 
double beta decay experiments.
The constraints we will derive \cite {mina-hiro02} are the 
maximal ones that can be derived in a model-independent way 
(and assuming ignorance of the CP violating phases) 
in a general framework of three-flavor mixing scheme of neutrinos. 
It utilizes the bounds imposed by neutrinoless 
double beta decay search and the CHOOZ reactor experiment 
\cite{CHOOZ}.

A natural question would be
"what is the impact of the KamLAND data on the double beta decay 
bound on neutrino mass and mixing parameters?".
The answer is that it allows us to derive, for the first time, 
an upper bound on the neutrino contribution to the cosmological $\Omega$ 
parameter, the energy density normalized by the critical density 
in the present universe. 
It is not because the parameter region for the LMA solution 
becomes narrower 
(in fact, the region for $\theta_{12}$ is virtually identical 
with that before KamLAND), 
but because the LOW and the VAC solutions which allow maximal
or almost maximal mixing are eliminated by the KamLAND data 
at 99.95~\% CL \cite{KamLAND}.

\section{Deriving the joint constraints by double beta decay and 
reactor experiments}

In this section we review our formalism presented in 
\cite{mina-hiro02}. 
We use the following standard parametrization 
of the MNS matrix~\cite{MNS}:
\begin{equation}
 U_{\mbox{\scriptsize{MNS}}}
  \equiv
  \left[
   \begin{array}{ccc}
    c_{12}c_{13} & s_{12}c_{13} & s_{13}e^{-i\delta}\nonumber\\
    -s_{12}c_{23}-c_{12}s_{23}s_{13}e^{i\delta} &
     c_{12}c_{23}-s_{12}s_{23}s_{13}e^{i\delta} & s_{23}c_{13}\nonumber\\
    s_{12}s_{23}-c_{12}c_{23}s_{13}e^{i\delta} &
     -c_{12}s_{23}-s_{12}c_{23}s_{13}e^{i\delta} & c_{23}c_{13}\nonumber\\
   \end{array}
  \right].
\label{MNSmatrix}
\end{equation}
 The mixing matrix for three Majorana neutrinos is
\begin{equation}
 U \equiv U_{\mbox{\scriptsize{MNS}}}
           \times \mbox{diag}(1, e^{i\beta}, e^{i\gamma}) ,
\end{equation}
where $\beta$ and $\gamma$ are extra CP-violating phases
which are characteristic to Majorana particles~\cite{Mphase}.
 In this parametrization, the observable of double beta decay experiments
is described as
\begin{eqnarray}
 \mbb
 &\equiv& 
  \left|\, \sum^{3}_{i=1} m_i U^2_{ei}\, \right| \nonumber\\
 &=&
  \left|\,
   m_1 c_{12}^2 c_{13}^2
   + m_2 s_{12}^2 c_{13}^2 e^{2 i \beta}
   + m_3 s_{13}^2 e^{2 i (\gamma - \delta)}\,
  \right| ,
\label{beta1}
\end{eqnarray}
where $U_{ei}$ denote the elements in the first row of $U$
and $m_i$ ($i=1,2,3$) are the neutrino mass eigenvalues.
 In the convention of this paper 
the normal hierarchy means $m_1 < m_2 < m_3$
and the inverted hierarchy $m_3 < m_1 < m_2$.

In order to utilize $\mbb^\ma$
which is an experimental upper bound on $\mbb$, 
we derive a theoretical lower bound on $\mbb$.
With appropriate choice of the phase-factor
$e^{2 i(\gamma - \delta)}$ in eq.~(\ref{beta1})
it leads to the inequality
\begin{eqnarray}
 \mbb^\ma
 \ge \mbb
 \ge c_{13}^2
         \left|
	  m_1 c_{12}^2 + m_2 s_{12}^2 e^{2 i \beta}
         \right|
       - m_3 s_{13}^2 .
\label{beta2}
\end{eqnarray}
Strictly speaking,  
the right-hand-side (RHS) of (\ref{beta2}) should be understood as 
the absolute value, but one can show that the alternative case 
does not lead to useful bound \cite{mina-hiro02}.
The RHS of (\ref{beta2}) is minimized by replacing  
$e^{2 i \beta}$ with $-1$ and $s_{13}^2$ with $s_\CH^2$, 
respectively, where $s_\CH^2$ is the largest value of $s_{13}^2$ 
which is allowed by the CHOOZ reactor experiment~\cite{CHOOZ}. 
Roughly speaking, $s_\CH^2 \simeq 0.03$.  
Thus, we obtain
\begin{eqnarray}
 \mbb^\ma
 &\ge& c_\CH^2 \left| m_1 c_{12}^2 - m_2 s_{12}^2 \right|
      - m_3 s_\CH^2\nonumber\\[3mm]
 &=& c_\CH^2
      \left\vert\,
       \frac{1}{\,2\,} (m_2 + m_1) \cos{2\theta_{12}}
       - \frac{1}{\,2\,} (m_2 - m_1)\,
      \right\vert
     - m_3 s_\CH^2.
 \label{beta3}
\end{eqnarray}
It is convenient to utilize $\cos{2\theta_{12}}$ as a parameter
because it arises naturally in the degenerate mass case.

Next, we derive a theoretical upper bound on $\mbb$
to utilize an experimental lower bound $\mbb^\mi$. 
Since the RHS of (\ref{beta1})
is maximized by setting the phase-factors to the unity,
we obtain
\begin{eqnarray}
 \mbb^\mi
  \le \left(m_1 c_{12}^2 + m_2 s_{12}^2 \right) c_{13}^2
        + m_3 s_{13}^2 .
\label{beta4}
\end{eqnarray}
Furthermore, the RHS of (\ref{beta4}) is maximized by 
replacing $s_{13}^2$ with $s_\CH^2$, and zero 
for the normal and the inverted hierarchies, respectively.
Then, we obtain 
\begin{eqnarray}
 \mbb^\mi
 &\le& \left(m_1 c_{12}^2 + m_2 s_{12}^2 \right) c_\CH^2
        + m_3 s_\CH^2\nonumber\\[3mm]
 &=& \left\{
      \frac{1}{\,2\,} (m_2 + m_1)
      - \frac{1}{\,2\,} (m_2 - m_1) \cos{2\theta_{12}}
     \right\} c_\CH^2
     + m_3 s_\CH^2
\label{beta5}
\end{eqnarray}
for the normal hierarchy,
and
\begin{eqnarray}
 \mbb^\mi
 &\le& m_1 c_{12}^2 + m_2 s_{12}^2\nonumber\\[3mm]
 &=& \frac{1}{\,2\,} (m_2 + m_1)
      - \frac{1}{\,2\,} (m_2 - m_1) \cos{2\theta_{12}}
\label{beta6}
\end{eqnarray}
for the inverted hierarchy.

The obtained bounds (\ref{beta3}), (\ref{beta5}), and (\ref{beta6}) 
can be represented as allowed regions on 
\mbox{$m_l$ - $\cos{2\theta_{12}}$} plane, 
where $m_l$ denotes the lightest neutrino mass for each hierarchy.
Notice that other mass eigenvalues are expressed by $m_l$ 
together with the solar and the atmospheric $\Delta m^2$'s;
\begin{eqnarray}
m_3^2 &=& m_1^2 + \Delta m^2_{12} + \Delta m^2_{23} ,
\nonumber \\
m_2^2 &=& m_1^2 + \Delta m^2_{12} ,
\label{m_i}
\end{eqnarray}
where $\Delta m^2_{ij} \equiv m^2_{j}- m^2_{i}$.

Alternatively, one can use $m_h$, the highest neutrino mass 
($m_3$ for the normal and $m_2$ for the inverted mass hierarchies) 
to make the similar plot, as done in \cite{mina-hiro02}. 
Whichever variable is used, they are observables in single beta 
decay experiments; modulo definitions of effective mass parameters 
\cite{mdef1,mdef2} 
it can be related, using (\ref{m_i}), the observable in direct mass 
measurement in tritium beta decay experiments. For a review, see 
e.g., \cite{beta}. 
We emphasize that importance of combining informations obtained 
by the single and the double beta decay experiments. 

We do not present a careful treatment of the uncertainty of 
nuclear matrix elements in this letter, but a brief comment 
will be made on how it affects the mass bound to be obtained 
in the next section.

\section{Analyzing the joint constraints by double beta decay and 
reactor experiments with presently available data}

\begin{figure}[h]
\begin{center}
\hspace*{-30mm}
\includegraphics[scale=0.35]{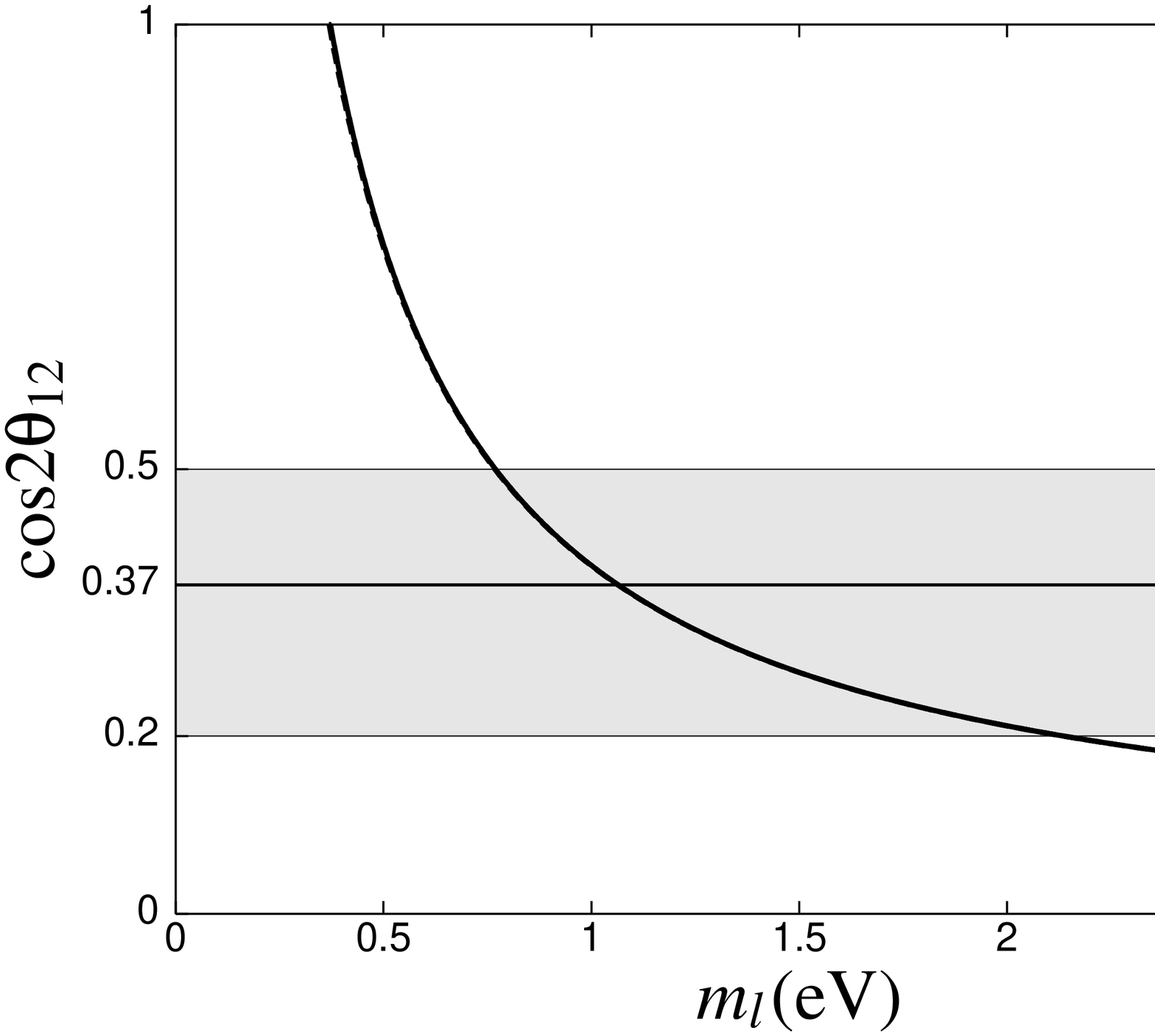}
\end{center}
\caption{
The double beta decay constraint with the CHOOZ bound 
(\ref{beta3}) is displayed 
as an allowed region in $m_l$ - $\cos{2 \theta_{12}}$ space. 
The region left to the curve is allowed. 
The solid and the dashed lines are for the normal and the 
inverted mass hierarchies, but they completely overlap.
The allowed LMA parameter region by the combined solar-KamLAND 
analysis is superimposed with shadow,
and $\cos{2\theta_{12}} = 0.37$ is the best fit value.
}
\label{presentbd}
\end{figure}

Let us first discuss the constraint based on the presently available 
experimental upper bound $\mbb < 0.35\,\eV$ \mbox{(90\,\% CL)} 
\cite{HeidelMoscow}. 
In Fig.~\ref{presentbd} we display the constraint imposed on 
mixing parameters on \mbox{$m_l$ - $\cos{2\theta_{12}}$} plane, 
where $m_l$ indicates the lowest mass, $m_1$ and $m_3$ for 
the normal and the inverted mass hierarchies, respectively.
The region left to the curves is allowed by the inequality 
(\ref{beta3}).
The solid and the dashed lines are for the normal and the inverted 
mass hierarchies, respectively. 
One notices that the two curves completely overlap with each other 
at the present sensitivity of the experiment.
The mass squared difference parameters are taken as 
$\Delta m^2_\atm = 2.5 \times10^{-3}\,\eV^2$, the best fit 
value in the atmospheric neutrino analysis \cite {shiozawa}, 
and $\Delta m^2_{12} = 7.3\times10^{-5}\,\eV^2$, 
the best fit value of the combined analysis of the KamLAND and 
all the solar neutrino data \cite{KL-solar}.

Also shown in Fig.~\ref{presentbd} as the shaded strip is the 
region allowed by joint analysis of the KamLAND and all the 
solar neutrino data \cite{KL-solar}. 
From  Fig.~\ref{presentbd} one can draw the upper bound on 
$m_l$ as $m_l \leq 2.13 (4.26)$ eV, where the number in parenthesis 
here and in eq.~(\ref{omega}) indicates the one for the case where 
a factor of 2 uncertainty 
of nuclear matrix elements is taken into account.
We notice that if either the LOW or the VAC solution were the 
solution to the solar neutrino problem the allowed region of 
$\cos{2\theta_{12}}$ would extend to zero (maximal mixing) and hence 
there would be no upper bound on $m_l$.

\begin{figure}[htbp]
\begin{center}
\vspace*{-20mm}
\hspace*{-30mm}
\includegraphics[scale=0.35]{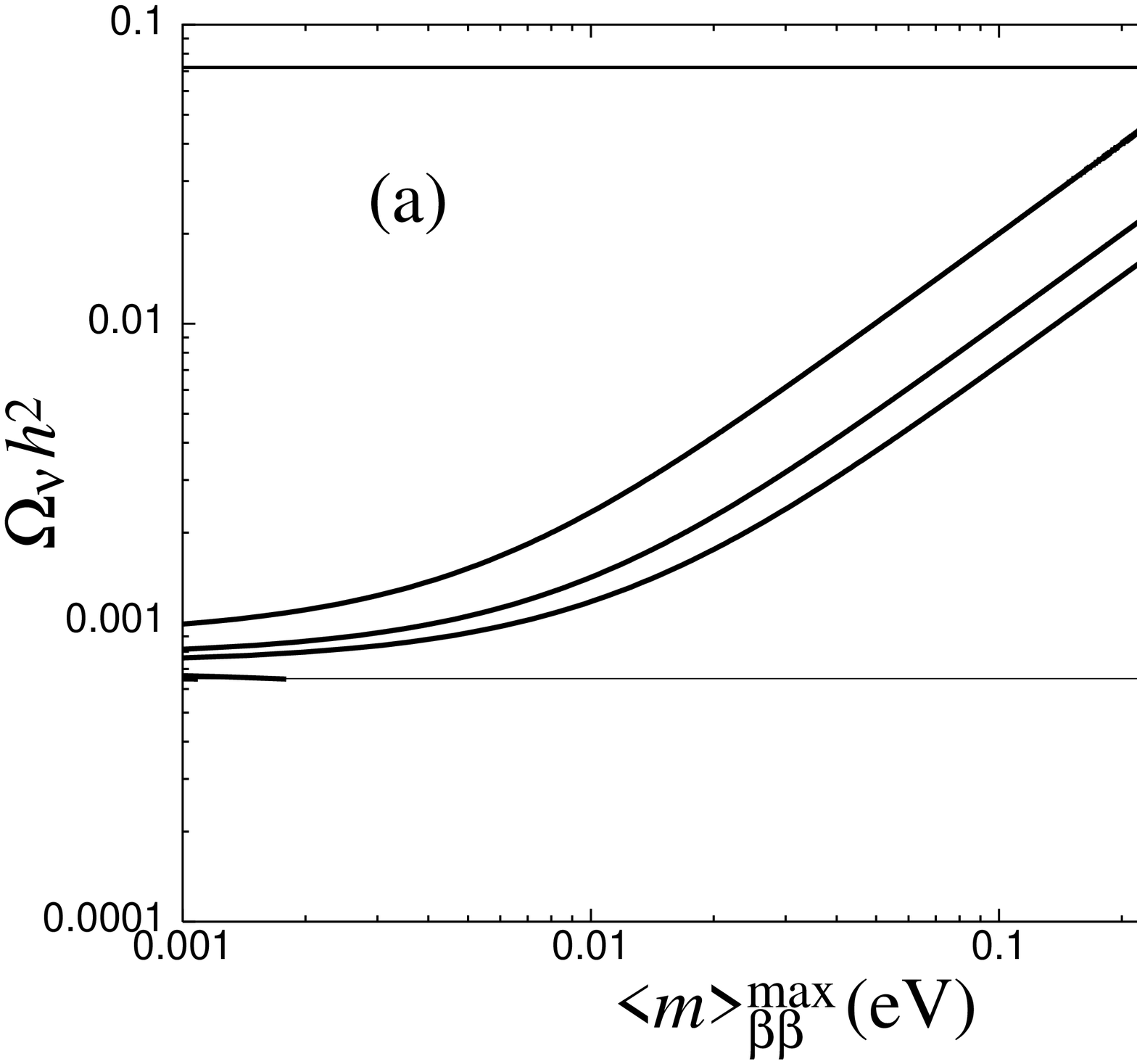}\\
\hspace*{-30mm}
\includegraphics[scale=0.35]{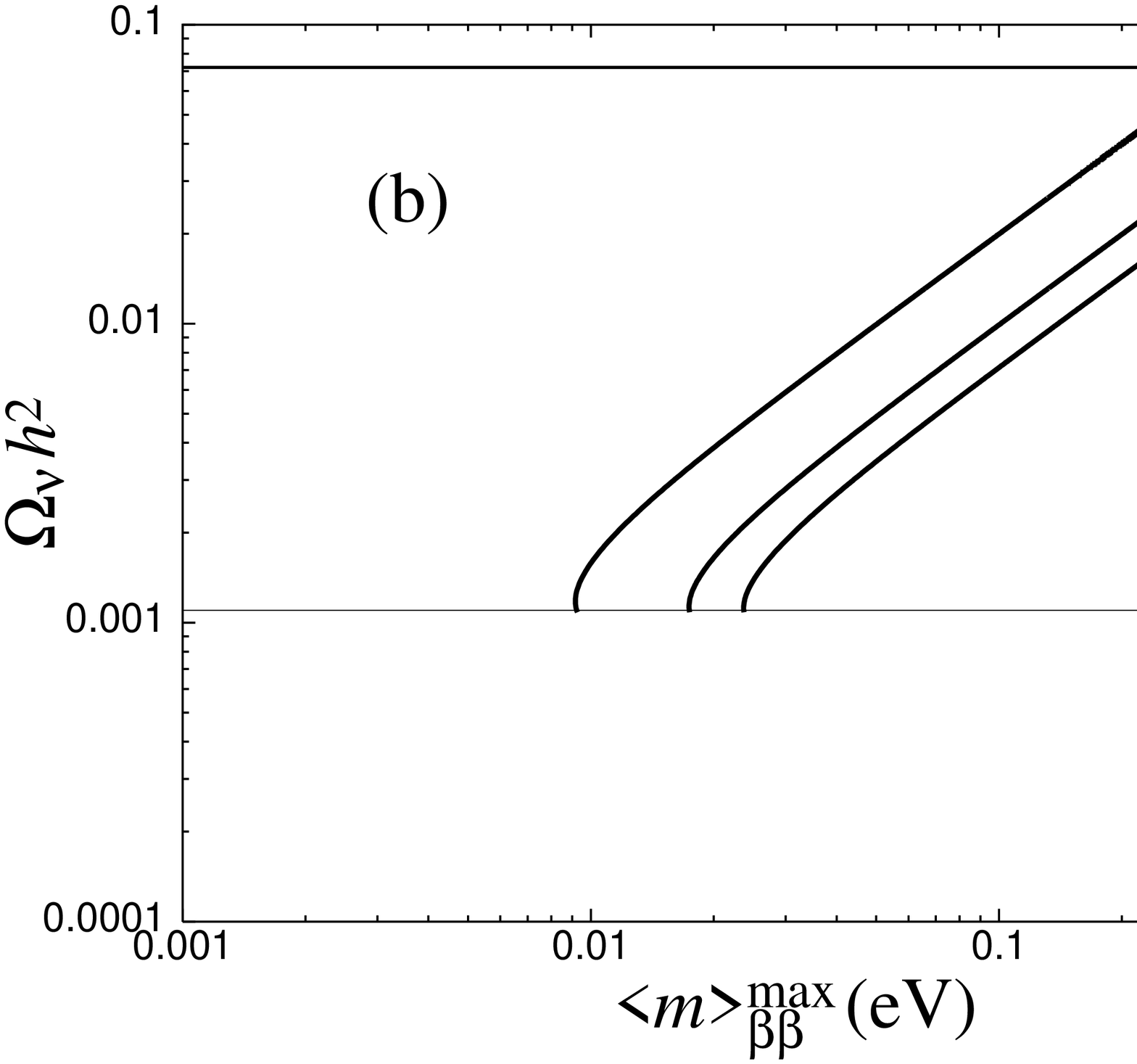}
\end{center}
\caption{
Plotted is the upper bound on $\Omega_{\nu}$, 
the neutrino contribution to cosmological $\Omega$ parameter.
Three lines correspond to, from above to below,  
$\tan^2{\theta_{12}} = 0.67,~0.46$, and 0.33, respectively.
Fig.~\ref{futurebd}a and \ref{futurebd}b are for the normal 
and the inverted mass hierarchies, respectively;
These correspond to the edges of the LMA region and the best fit value.
The bold horizontal line at $\Omega_{\nu} h^2 = 0.072$ 
is the upper bound determined by
the recent result of single beta decay experiments.
The thin horizontal line is the lower bound obtained by
$m_l = 0$.
}
\label{futurebd}
\end{figure}

This bound can be translated into the neutrino contribution to 
the cosmological $\Omega$ parameter \cite{Kolb}
\begin{eqnarray}
\Omega_{\nu} = 
\frac{\sum_i m_i}{91.5 \mbox{eV}} h^{-2} \leq 0.070 (0.14) h^{-2}
\label{omega}
\end{eqnarray}
which is already very restrictive. 
Here, $h$ denotes the Hubble parameter in units of 
$100 \mbox {km/{s Mpc}}$. 
For possible future use, we provide in Fig.~\ref{futurebd} 
the upper limit of $\Omega_{\nu}$ as a function of 
$\mbb^\ma$ for three values of $\tan^2{\theta_{12}}$, 
$\tan^2{\theta_{12}} = 0.46,~0.33$, and 0.67. 
These values are chosen as the best fit, the upper and the lower 
limit of 90 \% CL allowed region of the LMA solution by 
combined analysis of KamLAND and all the solar neutrino data 
\cite{KL-solar}.
It is notable that even the double beta bound 
$\mbb^\ma \leq 0.1$ eV which may be reachable in several years 
is powerful enough to constrain $\Omega_{\nu}$ parameter 
to be less than $\Omega_{\nu} \simeq 1 h^{-2}$~\%.

We notice that a short (almost) straight line, which is 
the lower bound on  $\Omega_{\nu} h^2$, appears at 
around the lower end in Fig.~\ref{futurebd}a. It is due to 
the fact that a vertical band appears in allowed region 
in $m_l$ - $\mbb$ plane at low $\mbb$, the feature first 
observed in \cite {mbbmin}. 
The behavior of the curve at around the lower end in  
Fig.~\ref{futurebd}b for the case of inverted mass hierarchy is 
slightly different from that obtained in \cite{barger's}.

We mention here that a similar constraint on neutrino 
masses may be obtained by cosmological observation 
\cite{cosmology}.

The constraints imposed on neutrino mixing parameters by 
neutrinoless double beta decay have been discussed by 
many authors. In a previous paper \cite{mina-hiro02} 
we tried to classify them into the ones in 
the "early epoch" \cite{early}, 
the "modern era" \cite {dblbeta1}, and 
the "post-modern era" \cite {dblbeta2} 
which now include some additional references.

\section{Analyzing the joint constraints by double beta decay and 
reactor experiments with possible future data}

In this section, we analyze the constrains based on the 
double beta decay bound which would be obtained in the future. 
We consider two sample cases. 

\vskip 0.4cm

\noindent
Case 1: $0.1\,\eV \le \mbb \le 0.3\,\eV$

\vskip 0.3cm

\noindent
Case 2: $0.01\,\eV \le \mbb \le 0.03\,\eV$.

\vskip 0.4cm

\noindent
Case 1 is a typical mass region proved and to be proved 
by the ongoing experiments \cite {HeidelMoscow,IGEX,NEMO}, 
while Case 2 is a typical one for the next-generation experiments. 
The latter include, in alphabetical order, 
CUORE, EXO, GENIUS, Majorana, MOON, and XMASS
\cite{CUORE,EXO,GENIUS,Majorana,MOON,XMASS}.
The mass square difference parameters are fixed as in the 
previous section.

\subsection{Case 1 : $0.1\,\eV \le \mbb \le 0.3\,\eV$}

\begin{figure}[htbp]
\begin{center}
\vspace*{-20mm}
\hspace*{-30mm}
\includegraphics[scale=0.35]{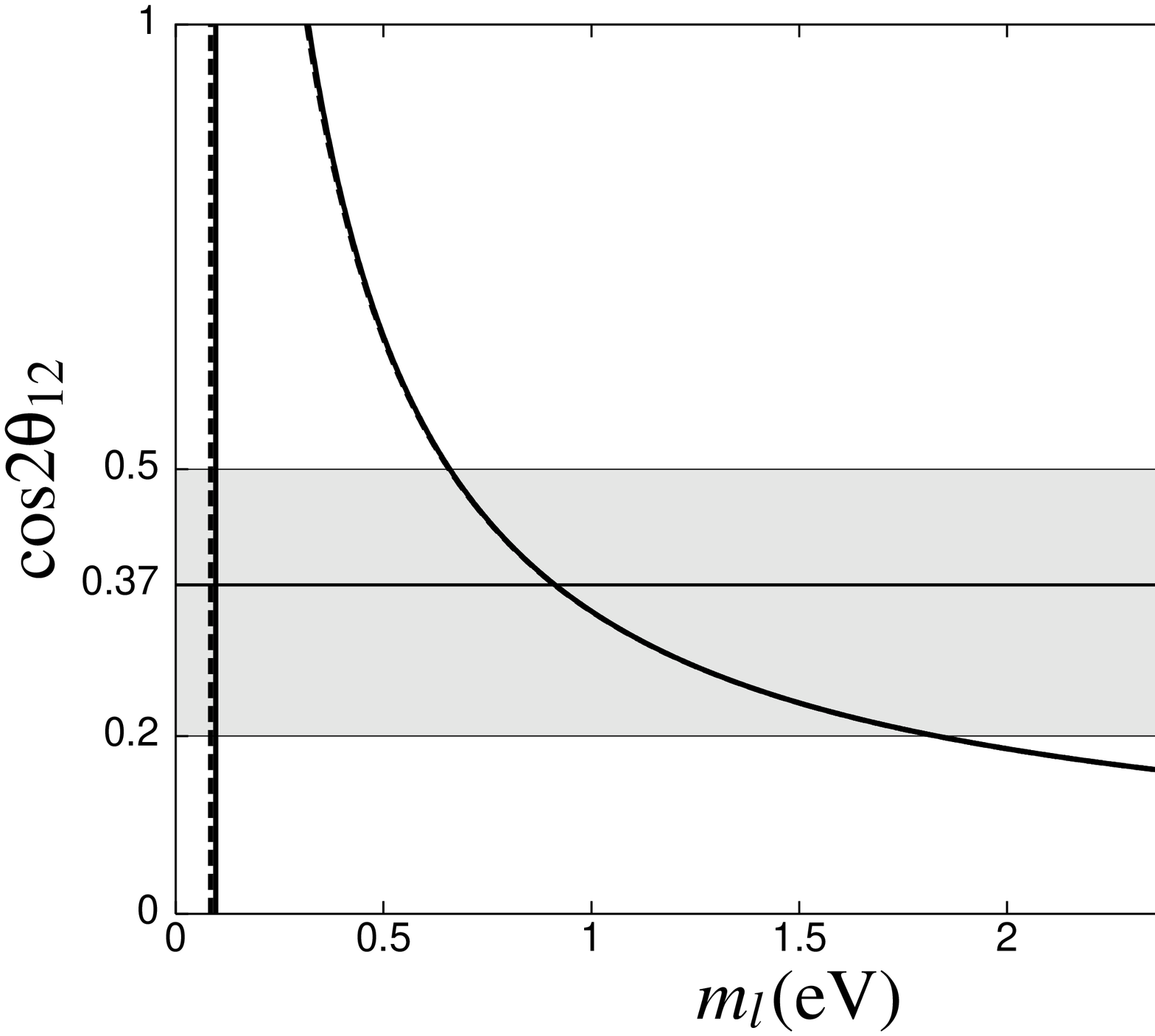}
\end{center}
\caption{
The double beta decay constraints with the CHOOZ bound 
(\ref{beta3}), (\ref{beta5}) and (\ref{beta6}) displayed for 
$0.1\,\eV \le \mbb \le 0.3\,\eV$. 
The solid and the dashed lines are for the normal and the 
inverted mass hierarchies, respectively.
The allowed LMA parameter region by the combined solar-KamLAND 
analysis is superimposed with shadow,
and $\cos{2\theta_{12}} = 0.37$ is the best fit value.
}
\label{case1}
\vspace*{-10mm}
\begin{center}
\hspace*{-30mm}
\includegraphics[scale=0.35]{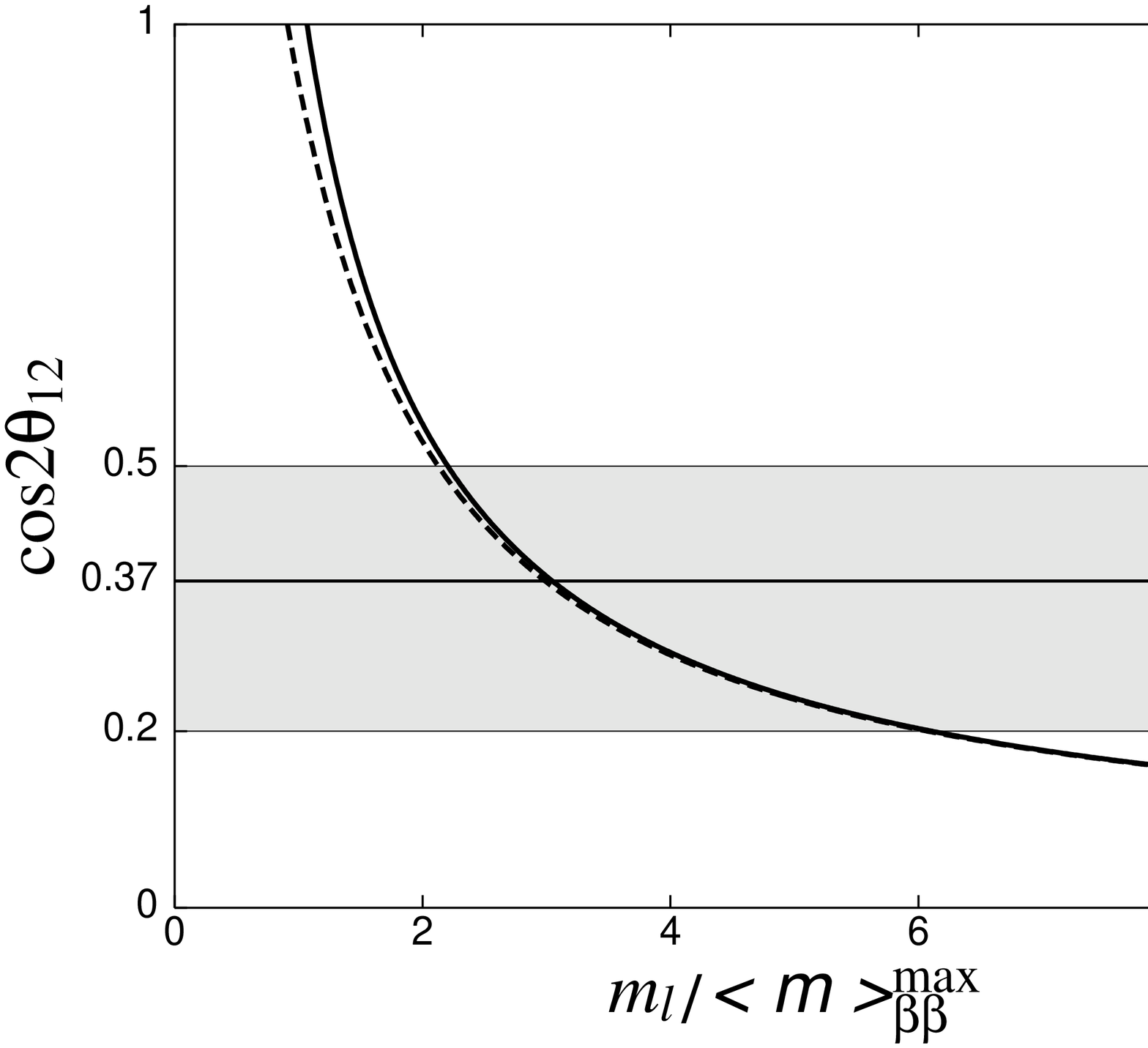}
\end{center}
\caption{
The double beta decay constraint with the CHOOZ bound 
(\ref{beta3}) is displayed for $\mbb^\ma = 0.1\,\eV$ to 
indicate the scaling behavior of the bound with variable 
$m_l/\mbb^\ma$ as used for abscissa. 
The region left to the curve is allowed. 
The solid and the dashed lines are for the normal and the 
inverted mass hierarchies, respectively.
The allowed LMA parameter region by the combined solar-KamLAND 
analysis is superimposed with shadow,
and $\cos{2\theta_{12}} = 0.37$ is the best fit value.
}
\label{degenerate}
\end{figure}

The bounds for this case are presented in Fig.~\ref{case1}. 
One notices that the lower bound on $\mbb$, 
(\ref{beta5}) and (\ref{beta6}), gives an almost 
vertical line at $m_l \simeq \mbb^\mi$ because of the small 
values of $\Delta m_{12}^2$.
Therefore, $\mbb^\mi$ is approximately regarded
as the lower bound on $m_l$.
In Case 2 with an order of magnitude smaller value of $\mbb$ 
the situation is quite different, as you see in the next subsection.

On the other hand, the upper bound on $\mbb$ gives a nontrivial 
constraint, as shown in Fig.~\ref{case1}. 
The curve for (\ref{beta3}) approaches to the asymptote 
$\cos{2\theta_{12}} = \pm t_\CH^2$ at large $m_l$, where 
$t_\CH = s_\CH / c_\CH$.
Hence, if either the LOW or the VAC solution were the case, 
there would be no upper bound on $m_l$. 
In the LMA solution the upper bound does exist 
for the current allowed region of $\cos{2\theta_{12}}$.

Notice that the bounds (\ref{beta3}) for the normal and the 
inverted hierarchies are very similar to each other.
It means that the degenerate mass approximation 
holds to a good accuracy.
In this approximation, the constraint (\ref{beta3}) becomes
\begin{equation}
 \mbb^\ma \ge m
               \left(
                c_\CH^2 \left| \cos{2\theta_{12}} \right| - s_\CH^2
	       \right) .
\end{equation}
where $m_i \simeq m$.

We now understand that in the degenerate mass regime  
$m_l/\mbb^\ma$ is a good parameter.
In Fig.~\ref{degenerate}, the bounds (\ref{beta3}) are presented
on the $m_l/\mbb^\ma$ - $\cos{2\theta_{12}}$ plane
for $\mbb^\ma = 0.1\,\eV$\@. 
Two curves for the normal and the inverted mass hierarchies 
overlap well, indicating goodness of the degenerate mass 
approximation.
Once we know that the scaling works, the result presented in 
Fig.~\ref{degenerate} applies to any values of $m_l$ and 
$\mbb^\ma$ as far as the degenerate mass approximation holds.

\subsection{Case 2 : $0.01\,\eV \le \mbb \le 0.03\,\eV$}

\begin{figure}[htbp]
\begin{center}
\hspace*{-30mm}
\includegraphics[scale=0.35]{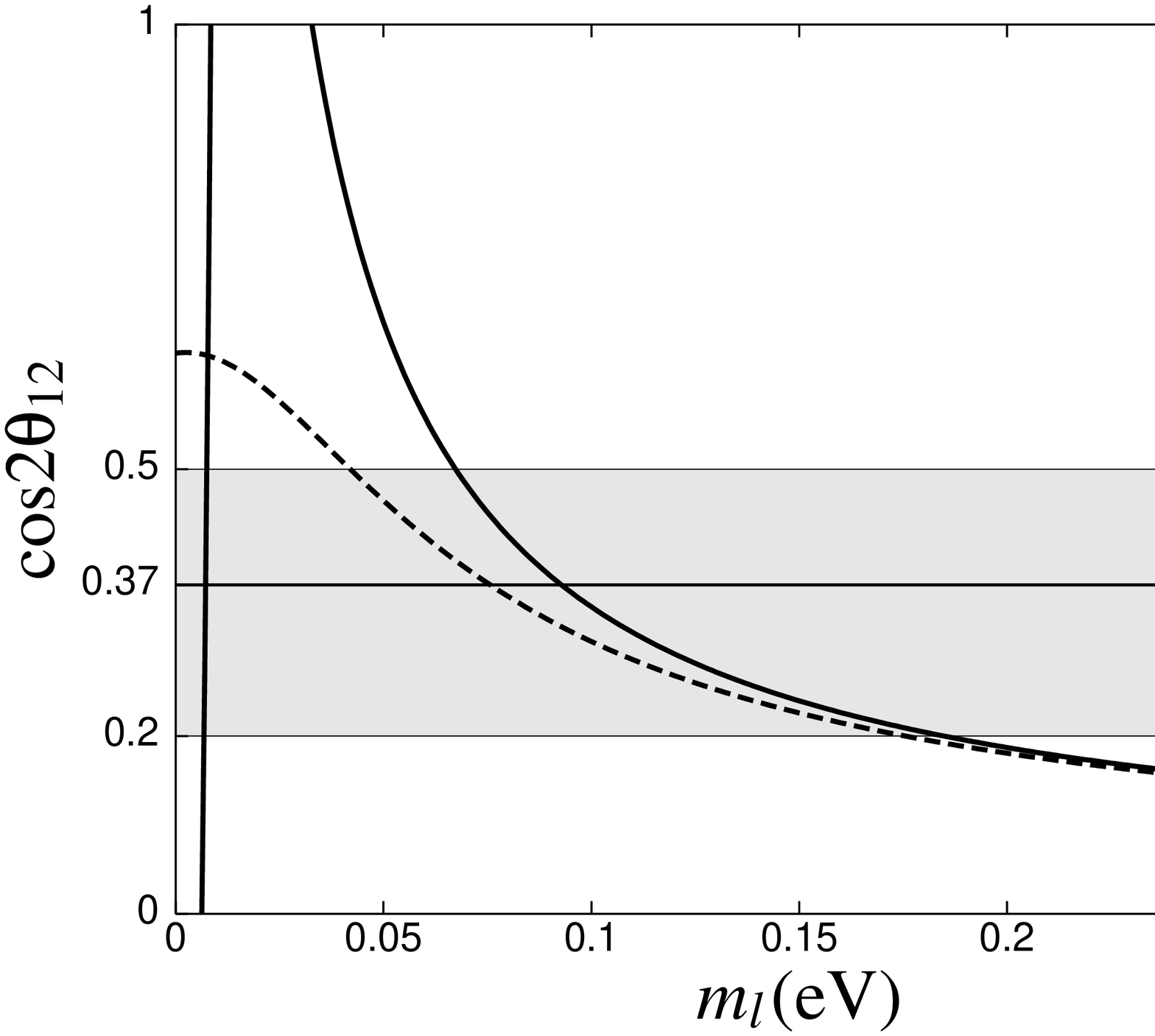}
\end{center}
\caption{
The double beta decay constraints with the CHOOZ bound 
(\ref{beta3}), (\ref{beta5}) and (\ref{beta6}) displayed for 
$0.01\,\eV \le \mbb \le 0.03\,\eV$. 
The solid and the dashed lines are for the normal and the 
inverted mass hierarchies, respectively.
The allowed LMA parameter region by the combined solar-KamLAND 
analysis is superimposed with shadow,
and $\cos{2\theta_{12}} = 0.37$ is the best fit value.
}
\label{case2}
\end{figure}

 The bounds for this case are presented in Fig.~\ref{case2}.
 It is clear that the degenerate mass approximation is no longer good
because the relevant energy scale is smaller than
the atmospheric mass scale $\sqrt{\Delta m^2_\atm} \simeq 0.05\,\eV$.%
\footnote{Although we put aside the uncertainties of
 $\Delta m^2$'s for simplicity,
 those affect particularly the values related to the lower bound on $m_l$
in such a small mass scale.
 Therefore, the values in (\ref{normal}), (\ref{inverted}),
and the left-hand-side of (\ref{byKDHK}) are just for reference.}
 The bounds for the normal and the inverted hierarchy significantly 
differ with each other.

An important point in Fig.~\ref{case2} is that with such improved 
sensitivities experiments may start distinguishing the two different 
mass hierarchies. Most notably, the difference between the 
normal and the inverted mass hierarchies is quite visible at 
$m_l \lsim 0.05$ eV, and the lower limit of $m_l$ 
disappears for the inverted hierarchy. 
It occurs because the constraint (\ref{beta6}) is satisfied
even for $m_l = 0$.
A similar things occurs also in the normal hierarchy if $\mbb^\mi$ 
is sufficiently small. 
One can easily estimate, by taking 
$m_l = 0$ and $\cos{2\theta_{12}} = 0$ in (\ref{beta5}) and (\ref{beta6}), 
the critical value of $\mbb^\mi$ under which the lower bound on 
$m_l$ disappears.
Approximately, they read 
\begin{equation}
 \frac{1}{\,2\,}\,c_\CH^2 \sqrt{\Delta m^2_{12}}
 + s_\CH^2 \sqrt{\Delta m^2_\atm}
 \simeq 0.0056\,\eV \sim \sqrt{\Delta m^2_{12}}
\label{normal}
\end{equation}
for the normal hierarchy,
and
\begin{equation}
 \frac{1}{\,2\,}\,c_\CH^2
  \left(
   \sqrt{\Delta m^2_\atm - \Delta m^2_{12}} + \sqrt{\Delta m^2_\atm}
  \right)
 \simeq 0.048\,\eV \simeq \sqrt{\Delta m^2_\atm}
\label{inverted}
\end{equation}
for the inverted hierarchy.
In Case 2, $\mbb^\mi = 0.01\,\eV$ is smaller than 0.048\,eV,
and that is why there is no lower bound on $m_l$ for the inverted hierarchy
in Fig.~\ref{case2}.

%


 Finally, let us extract the bound on $m_l$ from the result of 
\cite{evidence} by assuming that it will be confirmed by NEMO3.
Since $\mbb^\ma = 0.84\,\eV$ is large enough,
we can use Fig.~\ref{degenerate}.
Therefore, the upper bound on $m_l$ is extracted as
\begin{eqnarray}
 m_l \leq 3 [6]\times\mbb^\ma = 2.5 [5.0]\,\eV
\end{eqnarray}
for the LMA best fit parameters and for the most conservative case 
in square parenthesis, respectively.
%
%
On the other hand,
$\mbb^\mi = 0.05\,\eV$ determines the lower bound on $m_l$.
By combining those results, we obtain
\begin{eqnarray}
 0.05\,\eV (0.005\,\eV) \leq m_l \leq 2.5 [5.0]\,\eV
\label{byKDHK}
\end{eqnarray}
for the normal (inverted) hierarchy
with, in right-hand-side, 
the LMA best fit parameters and for the most conservative 
case in square parenthesis. 
%
%
Therefore, we see that all the active neutrinos is massive
if the result of \cite{evidence} is confirmed. This conclusion, 
however, may not survive if the uncertainties of $\Delta m^2$'s
and nuclear matrix elements are taken into account.

\section{Summary}

In the light of the KamLAND data we reanalyzed the constraint 
on neutrino masses and mixing imposed by neutrinoless double 
beta decay search and the CHOOZ reactor experiment.
We have shown that by now the absolute mass scale for neutrinos 
is bounded from above by the ``conspiracy'' of KamLAND and double 
beta decay experiments. This conclusion and our whole analysis 
in this paper are valid under the assumption that neutrinos are 
Majorana particles. 

In closer detail, allowed regions on a plane spanned by the 
lowest neutrino mass versus the solar mixing angle 
$\theta_{12}$ were obtained by using 
$\mbb^\ma$ ($\mbb^\mi$), the experimental upper (lower) bound 
obtained and to be obtained in the ongoing and the future 
double beta decay experiments.
For given $\theta_{12}$, these become constraints on 
neutrino mass, the lightest mass $m_l$ (or the heaviest 
one though we did not give them explicitly). 
Roughly speaking, 
$\mbb^\mi \leq m_l \leq 3 [6]\times\mbb^\ma$
for the LMA best fit parameters and for the most conservative case 
in square parenthesis, respectively.

It was made clear that
the condition $|\cos{2\theta_{12}}| > t_\CH^2 \simeq 0.03$
was necessary for the upper bound on $m_l$ to exist.
 On the other hand,
the condition $\mbb^\mi \geq 0.0056\,\eV (0.048\,\eV)$
needs to be satisfied
for the normal (inverted) mass hierarchy,
for the lower bound on $m_l$ to exist.
For example,
$0.05\,\eV \le \mbb \le 0.84\,\eV$
gives $0.05\,\eV (0.005\,\eV) \leq m_l \leq 2.5 [5.0]\,\eV$
for the normal (inverted) hierarchy.

\vskip0.5cm

\noindent
{\bf Note added}:

While this paper was in the revision process, we became aware of 
the report by the Wilkinson Microwave Anisotropy Probe (WMAP) \cite{WMAP}
in which a severe constraint on omega parameter 
$\Omega_{\nu} h^2 < 0.0076$ (95 \% CL) is placed.
It means that double beta decay searches must have the sensitivity 
in the region $\mbb \lsim 0.1\,\eV$ 
to be competitive, as one can see from Fig.~2.

\acknowledgments 

HM thank Theoretical Physics Department of Fermilab for 
hospitality where this work was completed. 
This work was supported by the Grant-in-Aid for Scientific Research 
in Priority Areas No. 12047222, Japan Ministry 
of Education, Culture, Sports, Science, and Technology.


\newpage

\newpage

\newpage

\newpage


\begin{thebibliography}{99}


\bibitem{KamLAND}
K.~Eguchi {\it et al.} [KamLAND Collaboration], 
Phys. Rev. Lett. {\bf 90} (2003) 021802 [hep-ex/0212021].


\bibitem{MSW}
S.~P.~Mikheev and A.~Y.~Smirnov,
Nuovo Cim.\ C {\bf 9} (1986) 17; \\ 
L.~Wolfenstein,
Phys.\ Rev.\ D {\bf 17} (1978) 2369.


\bibitem {solar}
Homestake Collaboration, K. Lande {\it et al.},
Astrophys. J.\ {\bf 496} (1998) 505;\\
%
SAGE Collaboration, J.\ N.\ Abdurashitov {\it et al.},
Phys.\ Rev.\ C {\bf 60} (1999) 055801; \\
%
GALLEX Collaboration, W.\ Hampel {\it et al.}, Phys.\
Lett.\  B {\bf447} (1999) 127; \\
%
Super-Kamiokande Collaboration,  S.\ Fukuda {\it et al.},
Phys. Rev. Lett. {\bf 86} (2001) 5651;
{\it ibid.}  {\bf 86} (2001) 5656;\\
%
SNO Collaboration, Q. R. Ahmad {\it et al.},
Phys. Rev. Lett. {\bf 87} (2001) 071301; 
 {\it ibid.} 
{\bf 89} (2002) 011301; {\bf 89} (2002) 011302. 


\bibitem {solaranalysis}
J.~N.~Bahcall, M.~C.~Gonzalez-Garcia and C.~Pena-Garay,
JHEP {\bf 0207} (2002) 054;\\
V.~Barger, D.~Marfatia, K.~Whisnant and B.~P.~Wood,
Phys.\ Lett.\ B {\bf 537} (2002) 179; \\
%
A.~Bandyopadhyay, S.~Choubey, S.~Goswami and D.~P.~Roy,
Phys.\ Lett.\ B {\bf 540} (2002) 14; \\ 
%
P.~C.~de Holanda and A.~Y.~Smirnov, hep-ph/0205241; \\
%
G.~L.~Fogli, E.~Lisi, A.~Marrone, D.~Montanino and A.~Palazzo,
Phys.\ Rev.\ D {\bf 66} (2002) 093008; \\
%
M.~Maltoni, T.~Schwetz, M.~A.~Tortola and J.~W.~Valle,
arXiv:hep-ph/0207227.



\bibitem {SKatm}
Kamiokande Collaboration, Y. Fukuda {\it et al.},
Phys. Lett. {\bf B335} (1994) 237;\\
Super-Kamiokande Collaboration, Y. Fukuda {\it et al.},
Phys. Rev. Lett. {\bf 81} (1998) 1562; 
{\it ibid.} {\bf 85} (2000) 3999.


\bibitem {K2K}
K2K Collaboration, S.~H.~Ahn {\it et al.},
Phys.\ Lett.\ B {\bf 511} (2001) 178;\\
K2K Collaboration, S.~H.~Ahn {\it et al.},
hep-ex/0212007.


\bibitem{MNS}
Z.~Maki, M.~Nakagawa and S.~Sakata,
Prog.\ Theor.\ Phys.\  {\bf 28} (1962) 870.
%



\bibitem{pontecorvo}
B.~Pontecorvo, Sov. Phys. JETP {\bf 26} (1968) 984.


\bibitem{vogel}
S.~R.~Elliot and P.~Vogel, hep-ph/0202264.



\bibitem {mina-hiro02}
H.~Minakata and H.~Sugiyama,
Phys.\ Lett.\ {\bf B532} (2002) 275 [hep-ph/0202003].


\bibitem {CHOOZ}
CHOOZ Collaboration, M.~Apollonio {\it et al.},
Phys.\ Lett.\ {\bf B420} (1998) 397;
{\it ibid.} B {\bf 466} (1999) 415. \\
See also, Palo Verde Collaboration,
F.~Boehm {\it et al.},
Phys.\ Rev.\ {\bf D64} (2001) 112001.


\bibitem{Mphase}
J.~Schechter and J.W.F.~Valle,
Phys.\ Rev.\ {\bf D22} (1980) 2227;\\
%
S.M.~Bilenky, J.~Hosek, and S.T.~Petcov,
Phys.\ Lett.\ {\bf B94} (1980) 495;\\
%
M.~Doi {\it et al.},
Phys. Lett. {\bf B102} (1981) 323.


\bibitem {mdef1}
Y. Farzan, O. L. G. Peres, and A. Yu. Smirnov, 
Nucl. Phys. {\bf B612} (2001) 59; \\
Y. Farzan and A. Yu. Smirnov, hep-ph/0211341.


\bibitem {mdef2}
C.~Weinheimer {\it et al.}, Phys. Lett. {\bf B460} (1999) 219; \\ 
F.~Vissani, Nucl.\ Phys.\ Proc.\ Suppl.\  {\bf 100} (2001) 273.


\bibitem{beta}
R.G.H.~Robertson and D.A.~Knapp, 
Ann.\ Rev.\ Nucl.\ Part.\ Sci.\ {\bf 38} (1988) 185.


\bibitem {HeidelMoscow}
Heidelberg-Moscow Collaboration, 
H.V.~Klapdor-Kleingrothaus {\it et al.}, 
Eur.\ Phys.\ J. {\bf A12} (2001) 147.


\bibitem {shiozawa}
M.~Shiozawa, Talk presented at XXth International Conference 
on Neutrino Physics and Astrophysics (Neutrino 2002), 
May 25-30, 2002, Munich, Germany.


\bibitem {KL-solar}
We use the value obtained by 
G.~L.~Fogli, E.~Lisi, A.~Marrone, D.~Montanino, A.~Palazzo, and 
A.~M.~Rotunno, hep-ph/0212127. \\
For the similar complehensive analyses, see \\ 
J.~N.~Bahcall, M.~C.~Gonzalez-Garcia and C.~Pena-Garay,
hep-ph/0212147; \\
%
M.~Maltoni, T.~Schwetz, and J.~W.~Valle,
hep-ph/0212129; \\
%
A.~Bandyopadhyay, S.~Choubey, R.~Gandhi, S.~Goswami and D.~P.~Roy,
hep-ph/0212146; \\
%
H.~Nunokawa, W.~I.~C.~Teves, and R.~Zukanovich Funchal, hep-ph/0212202; \\
%
P.~C.~de Holanda and A.~Y.~Smirnov, hep-ph/0212270; \\
%
V.~Barger and D.~Marfatia,  
hep-ph/0212126; \\
%
P.~Creminelli, G.~Signorelli, and A.~Strumia, 
hep-ph/0102234 (updated version).


\bibitem {Kolb}
E. W. Kolb and M. S. Turner, {\it The Early Universe} (Addison-Wesley
Publishing Co., California, 1990).


\bibitem {mbbmin}
H.V.~Klapdor-Kleingrothaus, H.~P\"as and A.Yu.~Smirnov, 
Phys. Rev. {\bf D63} (2001) 073005.


\bibitem {barger's}
V.~Barger {\it et al.}, Phys. Lett. {\bf B532} (2002) 15.


\bibitem{cosmology}
J.~Primack, astro-ph/0112255; \\
%
M.~Fukugita, G.-C.~Liu, and N.~Sugiyama, 
Phys.\ Rev.\ Lett.\ {\bf 84} (2000) 1082; \\
%
O.~Elgaroy {\it et al.}, Phys.\ Rev.\ Lett.\  {\bf 89} (2002) 061301.


\bibitem {early}
See, e.g., 
S.~T.~Petcov and A.~Yu.~Smirnov, Phys. Lett. {\bf B322} (1994) 109;\\
H.~Minakata, Phys. Rev. {\bf D52} (1995) 6630;
Phys. Lett. {\bf B356} (1995) 61; \\
S.~M.~Bilenky {\it et al.} Phys. Rev. {\bf D54} (1996) 4432. 


\bibitem {dblbeta1}
H. Minakata and O. Yasuda, Phys. Rev. {\bf D56} (1997) 1692; 
Nucl. Phys. {\bf B523} (1998) 597; \\
O. Yasuda, in {\it Proceedings of 2nd International Conference on 
Physics Beyond the Standard Model}, edited by 
H.~V.~Klapdor-Kleingrothaus and I. Krivosheina, 
pp 223-235 (IOP Bristol 2000); \\
%
T. Fukuyama, K. Matsuda, and H. Nishiura, 
Mod. Phys. Lett. {\bf A13} (1998) 2279;
Phys. Rev. {\bf D57} (1998) 5844; 
{\it ibid.} {\bf D62} (2000) 093001; {\bf D63} (2000) 077301; 
{\bf D64} (2001) 013001; \\
F. Vissani, JHEP {\bf 9906} (1999) 022; \\
V. Barger and K. Whisnant Phys. Lett. {\bf B456} (1999) 194; \\
J. Ellis and S. Lola, Phys. Lett. {\bf B458} (1999) 310; \\
G. C. Branco, M. N. Rebelo, and J. I. Silva-Marcos, 
Phys. Rev. Lett. {\bf 82} (1999) 683; \\
S. M. Bilenky, C. Giunti, W. Grimus, B. Kayser, and S. T. Petcov, 
Phys. Lett. {\bf B465} (1999) 193; \\
M. Czakon, J. Gluza, and M. Zralek, Phys. Lett. {\bf B465} (1999) 211; \\
H. Georgi and S. L. Glashow, Phys. Rev. {\bf D61} (2000) 097301; \\
R. Adhikari and G. Rajasekaran, Phys. Rev. {\bf D61} (2000) 031301; \\
D. Falcone and F. Tramontano, Phys. Rev. {\bf D64} (2001) 077302; \\ 
W.~Rodejohann, Nucl. Phys. {\bf B597} (2001) 110.


\bibitem {dblbeta2}
H.V.~Klapdor-Kleingrothaus, H.~P\"as and A.Yu.~Smirnov, 
cited in \cite{mbbmin}, \\
S. M. Bilenky, S. Pascoli, and S. T. Petcov, 
Phys. Rev. {\bf D64} (2001) 053010; \\
S. Pascoli, S. T. Petcov, and L. Wolfenstein, 
Phys. Lett. {\bf B524} (2002) 319; \\
%
H.~Minakata and H.~Sugiyama,
Phys.\ Lett.\ {\bf B526} (2002) 335; {\bf B532} (2002) 275 
[hep-ph/0111269, hep-ph/0202003]; \\ 
%
F.~Feruglio, A.~Strumia, and F.~Vissani, Nucl.\ Phys.\ {\bf B637} (2002) 345  
[hep-ph/0201291]; \\
%
W.~Rodejohann, hep-ph/0203214; \\
%
V.~Barger, S.~L.~Glashow, P.~Langacker, and D.~Marfatia, 
Phys. Lett. {\bf B540} (2002) 247; \\
%
H.~Nunokawa, W.~I.~C.~Teves, and R.~Zukanovich Funchal, hep-ph/0206137; \\
%
S. Pascoli and S. T. Petcov, 
Phys. Lett. {\bf B544} (2002) 239; \\
S. Pascoli, S. T. Petcov, and W.~Rodejohann, 
Phys. Lett. {\bf B549} (2002) 177; hep-ph/0212113. 


\bibitem{IGEX}
IGEX Collabolation,
C.E.~Aalseth {\it et al.},
Phys.\ Rev.\ {\bf D65} (2002) 092007.

\bibitem{NEMO}
F.~Piquemal (for the NEMO Collabolation), hep-ex/0205006.

\bibitem{CUORE}
E.~Fiorini {\it et al.} Phys.\ Rep.\ {\bf 307} (1998) 309;\\
%
A.~Bettini, Nucl.\ Phys.\ Proc.\ Suppl.\ {\bf 100} (2001) 332.

\bibitem{EXO}
S.~Waldman, Talk at International Workshop on Technology
and Application of Xenon Detectors (Xenon01), ICRR, Kashiwa, Japan,
December 3-4, 2001.


\bibitem{GENIUS}
GENIUS Collaboration, 
H.V.~Klapdor-Kleingrothaus {\it et al.}, hep-ph/9910205.

\bibitem{Majorana}
Majorana Collabolation,
C.E.~Aalseth {\it et al.},
hep-ex/0201021.

\bibitem{MOON}
H.~Ejiri {\it et al.},
Phys.\ Rev.\ Lett.\ {\bf 85} (2000) 2919.

\bibitem{XMASS}
S.~Moriyama, Talk at International Workshop on Technology 
and Application of Xenon Detectors (Xenon01), ICRR, Kashiwa, Japan, 
December 3-4, 2001.

\bibitem {evidence}
H.~V.~Klapdor-Kleingrothaus, A.~Dietz, H.~L.~Harnay,
and I.~Krivosheina,
Mod.\ Phys.\ Lett.\ {\bf A16} (2001) 2409 
[hep-ph/0201231].


\bibitem {WMAP}
D.~N.~Spergel {\it et al.}, astro-ph/0302209.


\end{thebibliography}
\end{document}